# BlockSim: An Extensible Simulation Tool for Blockchain Systems

Maher Alharby[1,2]* and Aad van Moorsel[1]

[1] School of Computing, Newcastle University, Newcastle upon Tyne, United Kingdom, [2] Department of Computer Science, Taibah University, Medina, Saudi Arabia

Both in the design and deployment of blockchain solutions many performance-impacting configuration choices need to be made. We introduce BlockSim, a framework and software tool to build and simulate discrete-event dynamic systems models for blockchain systems. BlockSim is designed to support the analysis of a large variety of blockchains and blockchain deployments as well as a wide set of analysis questions. At the core of BlockSim is a Base Model, which contains the main model constructs common across various blockchain systems organized in three abstraction layers (network, consensus, and incentives layer). The Base Model is usable for a wide variety of blockchain systems and can be extended easily to include system or deployment particulars. The BlockSim software tool provides a simulator that implements the Base Model in Python. The paper describes the Base Model, the simulator implementation, and the application of BlockSim to Bitcoin, Ethereum and other consensus algorithms. We validate BlockSim simulation results by comparison with performance results from actual systems and from other studies in the literature. We close the paper by a BlockSim simulation study of the impact of uncle blocks rewards on mining decentralization, for a variety of blockchain configurations.

OPEN ACCESS

*Edited by:*
Katinka Wolter,
Fachbereich Mathematik und Informatik, Freie Universität Berlin, Germany

*Reviewed by:*
Sam Maximilian Werner,
Imperial College London, United Kingdom
Philipp Reinecke,
Cardiff University, United Kingdom

*\*Correspondence:*
Maher Alharby
m.w.r.alharby2@ncl.ac.uk

*Specialty section:*
This article was submitted to Financial Blockchain, a section of the journal Frontiers in Blockchain

**Received:** 17 March 2019
**Accepted:** 06 May 2020
**Published:** 09 June 2020

*Citation:*
Alharby M and van Moorsel A (2020) BlockSim: An Extensible Simulation Tool for Blockchain Systems. Front. Blockchain 3:28. doi: 10.3389/fbloc.2020.00028

## 1. INTRODUCTION

In the design as well as the deployment of blockchain solutions, many architectural, configuration and parameterization questions need to be considered. Since it is usually not feasible or practical to answer these questions using experimentation or trial-and-error, model-based simulation is required as an alternative. In this paper, we propose a discrete-event simulation framework called BlockSim (Alharby and Van Moorsel, 2019) to explore the effects of configuration, parameterization and design decisions on the behavior of blockchain systems.

BlockSim aims to provide simulation constructs that are intuitive, hide unnecessary detail and can be easily manipulated to be applied to a large set of blockchains design and deployment questions (related to performance, reliability, security or other properties of interest). That is, BlockSim has the following objectives:

1. Generality: we want to be able to use BlockSim for a large set of blockchain systems, configurations and design questions.
2. Extensibility: BlockSim should be easy to manipulate by a designer or analyst to study different types and aspects of blockchain systems.
3. Simplicity: the above two objectives should be met while making BlockSim easy to use, both for simulation studies and for extending it.

This paper expands on the short introduction of the BlockSim framework in Alharby and Van Moorsel (2019), and discusses all facets of the tool design, implementation and use.

At the core of BlockSim is a Base Model, which contains model constructs at three abstraction layers: the network layer, the consensus layer and the incentives layer (Van Moorsel et al., 2018). The network layer captures the blockchain's nodes and the underlying peer-to-peer protocol to exchange data between nodes. The consensus layer captures the algorithms and rules adopted to reach an agreement about the current state of the blockchain ledger. The incentives layer captures the economic incentive mechanisms adopted by a blockchain to issue and distribute rewards among the participating nodes.

The Base Model contains a number of functional blocks common across blockchains, that can be extended and configured as suited for the system and study of interest. The main functional blocks include Node, Transaction, Block, Consensus and Incentives, as we describe in section 3. These are then implemented through a number of Python modules, discussed in section 4, and complemented by modules (event, scheduler, statistics, etc.) that implement the simulation engine.

The public nature of permissionless blockchains provides for particularly powerful opportunities to validate the simulator. We validate the BlockSim simulation results by comparing against theoretical results (invariants such as block rate), against data from the existing public blockchain systems such as Ethereum and Bitcoin and against results from the literature. The BlockSim simulation results are within a statistically acceptable margin of the real-life or published results, as discussed in section 6. We also demonstrate the use of BlockSim for a simulation study that considers stale rate, throughput and mining decentralization, for a range of possible blockchain configurations (not all existing in real-life systems). Using BlockSim we can demonstrate that uncle inclusion (such as in Ethereum) is beneficial for mining decentralization.

The structure of the paper is as follows. Section 2 discusses an overview of blockchain and its underlying layers. Also, it discusses an overview of modeling and simulation. Section 3 discusses the core Base Model of BlockSim including the design objectives behind it. Section 4 presents the implementation of the Base Model. Section 5 presents the application of BlockSim to Bitcoin, Ethereum and other consensus protocols as case studies. Section 6 discusses the validation of BlockSim against actual systems and studies from the literature. Sections 7 and 8 show a BlockSim simulation study as well as the evaluation of BlockSim against the design objectives. Section 9 discusses the related work. Section 10 concludes the paper.

## 2. BACKGROUND
### 2.1. Blockchain Overview

A blockchain is a distributed ledger, with an aim to keep track of all transactions that ever occurred in the blockchain network. This ledger is replicated and distributed among the network's nodes. Such a ledger has two main purposes, to provide an *immutable* log of all transactions, and to make the transactions *transparent* (i.e., visible) to anyone inspecting or using the blockchain.

The technologically most intriguing type of blockchain is the public or permissionless blockchain. The main feature of permissionless blockchains is that the nodes that participate in maintaining the ledger do not need to be trusted or even be known to each other. That is, any user can join and participate in the network. Permissionless blockchains contain a cryptocurrency, to reward nodes for investing resources in maintaining the blockchain. The first and most popular permissionless blockchain system is Bitcoin (Nakamoto, 2008), which is a digital payment system that enables non-trusting entities to commit financial transactions. Other blockchains (e.g., Ethereum, Wood, 2014) have introduced the idea of smart contracts to support various distributed applications such as e-voting, health applications, etc.

The term blockchain comes from the fact that data about multiple transactions is grouped into blocks. Each block is uniquely identified by its cryptographic hash and each block is attached and linked to the one that came before it. This results in a chain of blocks. Once a block is generated and attached to the blockchain ledger, the transactions in that block cannot be modified by any node, since it would require the node to rewrite all subsequent blocks. This makes blockchain systems immutable and protected against double-spending attacks (Alharby and Van Moorsel, 2017).

Any participating node in a permissionless blockchain can generate a transaction and broadcast it in the network. Each node has a pool to keep pending incoming transactions (transactions that need to be executed). To generate and attach a new block to the blockchain ledger, a subset of the nodes (called miners) select several pending transactions from their pools, execute them and then create a new block containing those transactions. How and when blocks are generated depends on the consensus protocol adopted by the blockchain system (see section 2.2.2).

Once a miner has successfully created a block, it will then broadcast it to other nodes in the network. Upon receiving the block, each node validates the block's correctness and appends it to its ledger. If the majority of the nodes attach the block to their ledger and start building on top of it, the block will be confirmed and considered as part of the blockchain ledger. The miner of that block can then collect a reward for the block as well as the fees associated with its transactions as compensation for their efforts.

### 2.2. Blockchain in Layers

Blockchain systems can naturally be divided in three layers, the network, consensus and incentives layer, as depicted in **Figure 1**. We will utilize these layers to structure the BlockSim simulator and therefore here provide a system explanation in layers as well. The network layer captures the network's nodes and the underlying network protocol to distribute information between nodes. The consensus layer captures the algorithms and rules adopted to reach an agreement about the current state of the blockchain ledger. The incentives layer captures the economic mechanisms adopted by a blockchain to issue and distribute rewards among the participating nodes.

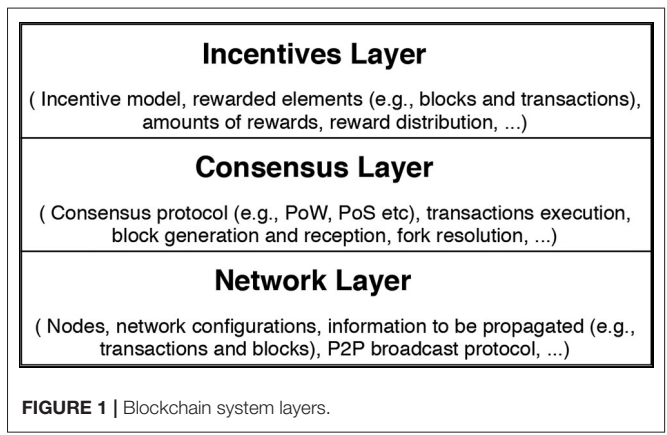

FIGURE 1 | Blockchain system layers.

## 2.2.1. Network Layer

The network layer in blockchain systems contains the nodes in the network, their geographical and relative locations and the connectivity among them. It defines which information is to be propagated as well as the mechanism to propagate such information.

The main constitute in the network layer is a *node*. A node can be an ordinary user who wants to create and submit a transaction to be executed and included in the ledger or a special node, known as *miner*, who maintains and expands the ledger by appending new blocks. A node has a unique identifier and maintains its balance, a local copy of the blockchain ledger and, if the node is a miner, an individual transactions pool. The transactions pool keeps the pending transactions received from other nodes in the network.

Nodes communicate the following information to each other. If a node generates a new transaction, it cryptographically signs it and propagates it to its peers to have it confirmed and recorded in the blockchain ledger. In case the node is a miner, every time it generates a block, it notifies its peers so they can validate it and append it to their copies of the ledger.

As information propagation mechanism for blockchains several protocols have been proposed, including relay networks and advertisement-based protocols (Gervais et al., 2016). In the advertisement-based protocol used in most blockchains (Gervais et al., 2016), the node sends a notification to its peers about the new data (e.g., a transaction). If the recipient node responds by requesting the data, the node will send it. Otherwise, the node will not send it as the recipient node has already had the data.

## 2.2.2. Consensus Layer

The consensus layer in blockchain systems defines the algorithms and rules for reaching an agreement about the blockchain's state among the network's nodes. Such rules specify which node is eligible for generating and appending the next block to the blockchain ledger, how often blocks are generated as well as how to resolve potential conflicts that may occur when nodes have multiple, differing copies of the ledger.

There are several consensus algorithms such as Proof of Work (PoW) and Proof of Stake (PoS) that have been proposed for blockchain systems. In PoW, nodes (i.e., miners) invest their computing power to maintain the ledger by attaching new blocks, while in PoS, nodes invest their stake or money. Regardless of what is required to be invested by the nodes, the intuition behind such algorithms is to introduce a cost for maintaining the ledger. The cost introduced should be more than enough to deter nodes from behaving maliciously (Wang et al., 2019). At the same time, nodes are only rewarded for their efforts if they follow the rules and maintain the ledger honestly (see section 2.2.3).

To illustrate the consensus layer, we discuss the PoW algorithm here as it is the most common algorithm for permissionless blockchains, used by Bitcoin and Ethereum. In PoW, the computing power invested by a miner determines how frequent that miner will generate and append blocks to the blockchain ledger. To generate a block, the miner has to repeatedly try nonces (random numbers) until the hash of the nonce combined with the block information will be within a certain threshold (referred to as the block difficulty). The only way to find the nonce is by trial-and-error, and thus, the more hash power invested by a miner, the more likely that miner will find the nonce. This process is a competitive task since all miners in the network are competing against each other to find the desirable hash value of the next block. Note that the block difficulty can be dynamically adjusted to control how often blocks are generated.

Due to the delay incurred by propagating blocks between nodes in the network (see Network Layer), other nodes might generate the next block before hearing of another competitive block that has recently announced. This leads to conflicts, known as forks, which occur when nodes have multiple, differing views of the ledger. The task of the consensus layer in blockchain systems is to resolve such conflicts. Different consensus algorithms use different rules to select which blockchain (fork) should be accepted as the global chain. For example, the PoW algorithm used by Bitcoin and Ethereum resolves the conflicts by adopting the longest chain. Other proposals such as GHOST (Sompolinsky and Zohar, 2015) select the fork with the heaviest work.

## 2.2.3. Incentives Layer

The incentives layer utilizes the blockchain's cryptocurrency to establish an incentive structure, distributing rewards among the participating miners who maintain the blockchain's ledger. The incentive model is essential to maintain any permissionless blockchain system. Incentives should compensate miners fairly for their work and motivate them to behave honestly (Aldweesh et al., 2018; Alharby et al., 2018). The incentives also protect the blockchain system from various attacks (e.g., DDoS attacks in Ethereum, Buterin, 2016) and against malicious behaviors of the nodes (e.g., selfish mining strategies, Eyal and Sirer, 2018).

In most blockchain systems rewards are associated with generating blocks and completing transactions, called block reward and transaction fees, respectively. Depending on the chain, there are subtle differences in what is rewarded, e.g., Ethereum issues a reward for stale (or uncle) blocks, even if they do not make it into the blockchain when conflicts are resolved. When a miner receives a reward (e.g., through appending a new block to the ledger), its balance will increase accordingly. The block reward, in all known blockchains, is set to a fixed amount, while the transaction fee is calculated as a variable amount of cryptocurrencies, depending on effort as well as the prize a transaction submitter is willing to pay.

## 2.3. Modeling and Simulation

A model is an abstract representation of a real system, either existing or in design. The model usually comprises mathematical expressions as well as structural and logical relationships to describe the dynamics of the system (Anderson et al., 2015). Simulation is a quantitative method, which "executes" the model to mimic the behavior of the system (Anderson et al., 2015). Simulation can be used to predict and describe how different configurations and scenarios impact the behavior of the system. Thus, simulation can be used to answer "what-if" questions and to experiment with new designs and policies without a need to interrupt a functioning system (Banks, 1984).

Simulation can be classified into two categories, namely, discrete-event simulation and continuous-event simulation (Haverkort, 1998). Human-made systems such as digital computer and information systems are most suitable represented as discrete-event simulation, as the systems change state at discrete moments in time (Fishman, 2001). BlockSim utilizes the discrete-event simulation approach to design and implement the simulator.

There are two approaches to develop simulation tools, namely, general-purpose programming languages (e.g., C++, Java, or Python) and special-purpose simulation languages (e.g., Arena and GPSS) (Leemis and Park, 2006). The former is more flexible and familiar, while the latter provides several built-in features (e.g., statistics, event scheduler, and animation) that reduce the time required to build models. As stated in Leemis and Park (2006), there is a debate and conflict about which method is preferable. Also worth noting are simulation frameworks that enable developing simulation models using general-purpose languages, for instance, OMNeT++ and SimPy for developing models in C++ and Python.

To develop and implement BlockSim, we opt for Python as a general-purpose language. We do not use its simulation framework *SimPy* because it follows a process-oriented paradigm, which differs from the approach we take in BlockSim. However, it will be useful to consider integrating features provided by SimPy with our simulator in a future version.

## 3. BLOCKSIM BASE MODEL

In this section, we introduce the *Base Model* underlying BlockSim, which is designed to model any kind of blockchain system, with application specific extensions as needed. We first define the design principles and goals for BlockSim: generality, extensibility and simplicity. Then, we discuss the design layer by layer: Network Layer, Consensus Layer, and Incentives Layer. Within each layer we identify the key functional units (entities) and the actions or activities it executes.

## 3.1. Design Principles

We design a Base Model to fulfill the main design goals for BlockSim, which are:

- **Generality:** we want to be able to use BlockSim for a large set of blockchain systems, configurations and design questions.
- **Extensibility:** BlockSim should be easily manipulated by a designer or analyst to study different aspects of blockchain systems.
- **Simplicity:** the above two objectives should be met while making BlockSim easy to use, both for simulation studies and for extending it.

The art of designing a tool such as BlockSim is to find a useful trade-off between generality and extensibility on the one hand, and simplicity to achieve these two objectives on the other hand. The Base Model is critical in achieving this goal, aiming to find the optimal trade-off among the above three objectives for the domain of blockchain systems.

The Base Model identifies the key building blocks (e.g., blocks, transactions, nodes, and incentives) common across all blockchains BlockSim is meant for, see **Figure 2**. The Base Model dictates how general the model class is that is supported by BlockSim, and particularly how easy it is to build new models. The Base Model will be translated in software modules and therefore also determines if BlockSim can be extended easily, for instance, to provide more detailed models of certain processes that take place in blockchains.

## 3.2. Network Layer

This layer defines two entities *Node* and the underlying *Broadcast protocol*, as depicted in **Figure 2**. The *Node* entity is responsible for updating the system state variables (e.g., the blockchain ledger and the transactions pool). The *Broadcast protocol* specifies how information entities (e.g., *Blocks* and *Transactions*) are propagated in the network.

Both *Blockchain ledger* and *Transactions pool* entities are part of the *Node* entity (see **Figure 2**). That is, every node maintains and continuously updates these entities. We model nodes as objects that have different attributes such as unique ID, balance, local ledger and transactions pool. The transactions pool and the local ledger are modeled as array lists that can be extended when new transactions and blocks are received. These attributes are common across the different implementation of blockchains. It could, however, be possible to extend this by including more additional attributes, as we will show in section 5.1.

The propagation of information entities depends on the *Broadcast protocol* entity, which can be modeled in detail by accounting for the network configurations, the geographical distribution of the nodes and the connectivity among the nodes, or it can be modeled in an abstraction level by only considering a time delay for propagating information among the nodes. The reason for abstracting the broadcast protocol is to make our simulator as simple as possible by hiding unnecessary details. This will alleviate the user of the simulator from configuring many parameters related to the network configurations such as the broadcast protocol, the geographical distribution of the nodes and the number of connections per node. Having the propagation delay as the only configurable parameter will improve both the efficiency and the usability aspects of the simulator.

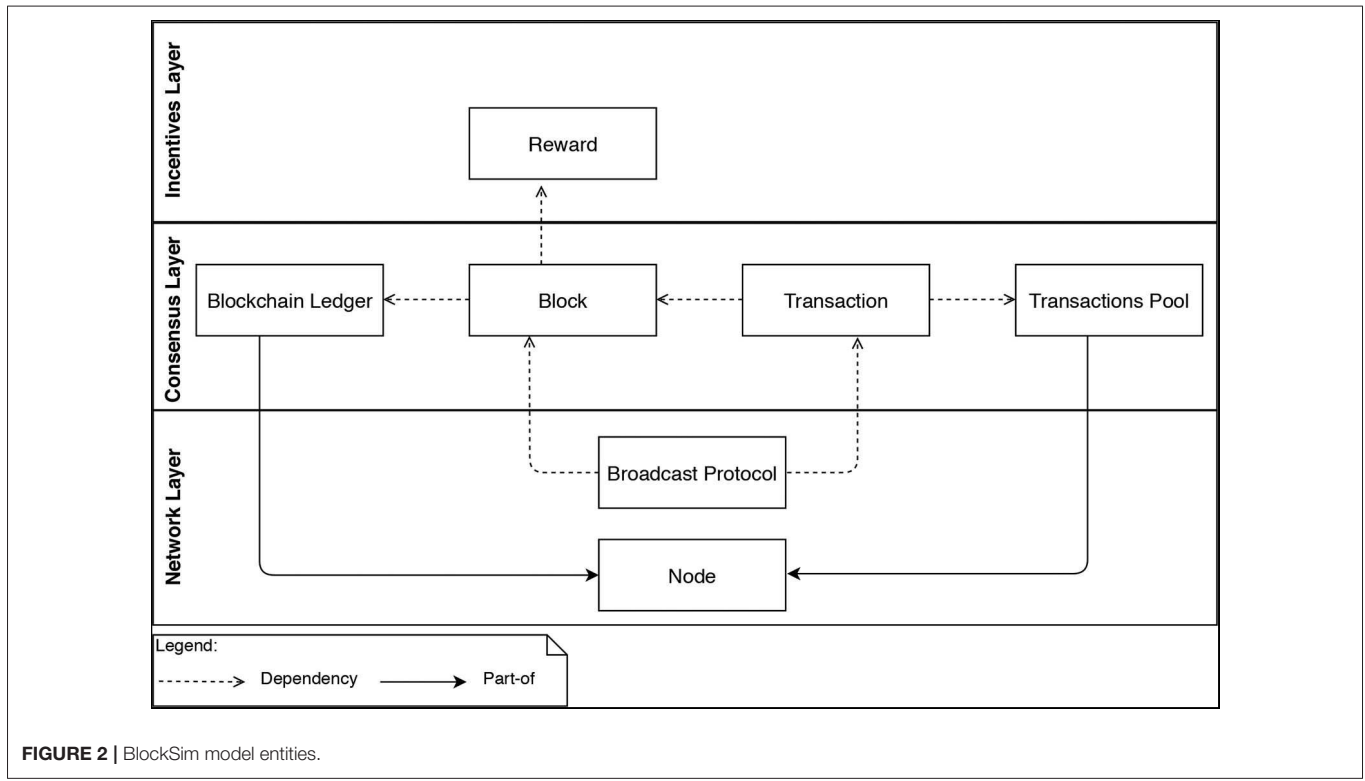

FIGURE 2 | BlockSim model entities.

## 3.3. Consensus Layer

This layer aims at establishing the rules that nodes can follow to reach an agreement about the blockchain's state. This layer includes four entities, namely, *Transaction*, *Block*, *Transactions pool*, and *Blockchain ledger*, as depicted in **Figure 2**.

The *Blockchain ledger* entity depends on the *Block* entity, and the *Block* entity depends on the *Transaction* entity. That is, the blockchain ledger is composed of blocks and blocks are composed of transactions. The *Transactions pool* depends on the *Transaction* entity, as every transaction created is fed into the transactions pool. The *Node* entity maintains these four entities.

Within the consensus layer, there are several activities or actions to be executed by the entities. The creation of blocks and transactions is an example of such activities. The flow of these activities is depicted in **Figure 3**. These activities run continuously, transactions and blocks, for instance, always keep arriving in the network.

### 3.3.1. Transaction

Transactions are one of the building blocks (entities) common across all blockchain systems. It plays a significant role in updating the blockchain's state. The arrival of a new transaction in the network results in updating the transactions pool by inserting that transaction.

We model transactions in two different ways, namely, full and light. The full technique helps to track each transaction in the system (e.g., when a transaction has been created and included in a valid block). This technique models transactions as in any blockchain system, and it is useful if one is interested in, for instance, studying the latency of individual transactions in blockchain systems. However, this type of modeling consumes an enormous amount of computing resources and time during the simulation since each transaction has to be tracked. On the other hand, the light technique does not track each transaction. It is useful when studying the throughput of blockchain systems without caring about the confirmation time of transactions within the system.

In both techniques, we model transactions as objects that have several attributes or fields such as transaction ID, size, fee, timestamp, contents as well as the submitter and the recipient of the transaction. These attributes are almost common across all blockchains, and that some systems have more additional attributes (e.g., Ethereum has also gas-related attributes such as Gas Limit).

**Full modeling Technique**: In this technique as we discussed in section 3.2, we model an individual transactions pool for each node by assigning an array list for each node as a way to abstract the pool. Each transaction created by a node is propagated to all other nodes in the network. Upon receiving the transaction, the recipient node appends it to their pool. Thus, we model transactions in three different activities labeled from 1 to 3, as depicted in **Figure 3**.

- **Creating transactions:** This involves generating transactions by the participating nodes. The number of transactions to be created per unit of time can be controlled and configured.

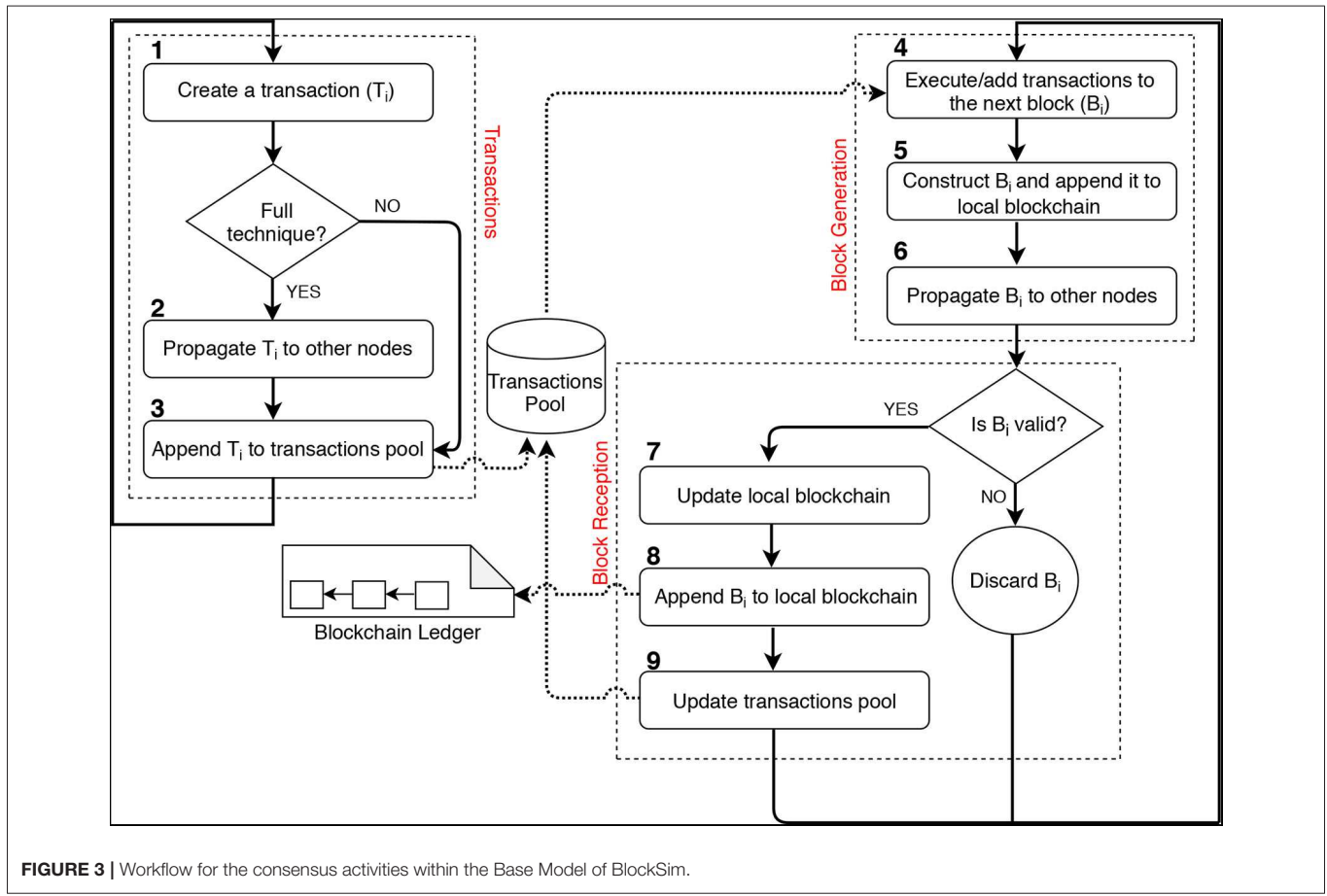

FIGURE 3 | Workflow for the consensus activities within the Base Model of BlockSim.

- **Propagating transactions:** This requires the creator of the transaction to propagate it to other participating nodes. This is to notify other nodes about the newly created transactions.
- **Appending transactions:** This requires the recipient of the transaction to append it to their transactions pool.

**Light modeling Technique**: In this technique, we only model a single transactions pool to be shared among all nodes in the network. The intention behind this technique is to provide an alternative and simplified way to model transactions by omitting the propagation process as well as the needs for nodes to update their pools continuously (see section 3.3.2). Thus, the light technique is more efficient and faster during the simulation. However, this technique cannot be used to draw conclusions about the latency of transactions as transactions are not tracked. Nevertheless, it is useful to get indicators about the throughput in blockchain systems.

In this technique, we create a set of transactions (N) and then append it to the shared pool before the mining process, so miners can access the pool to select several transactions to include in their forthcoming block. Hence, N should be more than enough for a block, usually enough for two blocks. Once a miner has successfully generated a block, the pool is reset and then filled up with a fresh set of transactions to be included in the next block.

Both techniques could be implemented and then the user would be given a choice to select which method to adopt based on their own needs. For instance, if one is only interested in throughput, there is no need for choosing the full technique since it makes the simulator runs for a very long time.

### 3.3.2. Block

Blocks are another essential building block (entity) of any blockchain system. Blocks consist of transactions. The arrival of a new block results in an update in the transactions pool and blockchain ledger. The pool is updated by removing all transactions included in the block, while the ledger is updated by appending the newly created block.

We model blocks as objects that have several attributes, namely, depth, block ID, previous block ID, timestamp, size, miner ID, and transactions. The block ID is a unique identifier for the block. The block depth indicates the index of the block in the node's blockchain. The miner ID refers to the node that created the block. Each block can accept a list of transactions as its content. These attributes are common across blockchains.

We model blocks in the consensus layer as *Block Generation* and *Block Reception*, see **Figure 3**. Block generation specifies when blocks are generated as well as which node is eligible for appending the next blocks. It covers all the common actions required by a miner to create and attach a block to the blockchain ledger. The actions embrace executing the block's transactions, constructing and appending the block to the local blockchain and propagating the block to other nodes in the network. Block reception specifies how the network's nodes update their blockchain ledgers upon receiving new blocks. It covers the common activities taken by a node when receiving a newly generated block. Upon receiving a valid block, the recipient node will perform three actions, which are updating local blockchain if necessary, appending the block to the local blockchain and updating the transactions pool.

The consensus algorithm is responsible for selecting a miner to build the next block. The methodology used to choose a miner varies among blockchains, depending on the adopted consensus protocol. In PoW, for instance, miners are selected based on solving a mathematical task. Once a miner is chosen to construct and append a new block to the ledger, the miner would undertake the following actions. Hence, these actions are common across all blockchain systems, and that some specific systems may include other activities (e.g., including uncle blocks in a future block as in Ethereum).

- **Executing and adding transactions to the block:** This requires the miner to select several pending transactions to be executed and included in the next block. Often, miners first sort those pending transactions based on their associated fees. Then, miners select the best transaction according to their ranking criteria, execute it if and only if there is a space in the block. The transaction will then be recorded in the block. After that, miners will select the next transaction and continue until the block is full or there is no pending transaction.
- **Constructing and appending the block to the local blockchain:** After preparing the block content (e.g., transactions), the miner would construct the block after which the block will be appended to the miner's local blockchain.
- **Propagating the block to other nodes:** This is to propagate the block to other nodes in the network. This is to notify the network's nodes about the newly generated block.

Once a node has received a new block, it will check its validity. The block is considered valid if it was constructed correctly and all embedded transactions were correctly executed. Beside the block validity, the block must point to the last block in the ledger (the block's depth should be higher than that of the last block). We only model the block depth, and thus, we abstract the validity of the block. If the depth of the received block is not higher than that of the last block, the block will be discarded. Otherwise, the node will perform the following actions.

- **Updating local blockchain:** This requires the recipient node to update its local blockchain, where necessary, before appending the newly received block. This is because sometimes the received block is built on different preceding blocks (a different chain branch) compared to the ones the recipient node has or because it is built on missing blocks. Therefore, the node has to update all the preceding blocks (and fetch all missing blocks if any) according to the ones the received block is following.
- **Appending the block to local blockchain:** This is to append the received block to the local copy of the blockchain.
- **Updating transactions pool:** This requires the recipient node to update its transactions pool, where necessary, upon appending the newly received block. This is to remove all the transactions that have already been executed in the received block from the node's pool.

### 3.3.3. Transactions Pool and Blockchain Ledger

Transactions pool and blockchain ledger are also important building blocks (entities) since they represent the state of blockchain systems. The transactions pool is updated upon the arrival of a new transaction or block, while the blockchain ledger is only updated once a block has arrived, as discussed in sections 3.3.1, 3.3.2. Nodes are responsible for updating both the pool and the ledger, as every node in the blockchain network maintains a local copy of them (see section 3.2).

**The rule of updating the ledger in the case of forks:** Nodes at some point in time may have different views of the blockchain ledger due to the network's propagation delay. A significant role of the consensus layer is to define the rules that can be used by the nodes to resolve the forks. For instance, Bitcoin and Ethereum use the longest-chain rule to resolve the forks. That is, nodes update their ledgers every time they receive a block that follows a chain that is longer than their local chains. By doing so, nodes will have the same view of the blockchain ledger. Other systems, however, use different rules (e.g., GHOST, Sompolinsky and Zohar, 2015).

## 3.4. Incentives Layer

The incentives layer is responsible for designing the underlying incentive model by defining the rewarded elements (e.g., blocks and transactions) as well as distributing the rewards among the participating miners. This layer has the *reward* entity, which depends on the *Block* entity (see **Figure 2**). That is, the rewards are only given to the miners upon appending new blocks to the ledger. The calculation and the distribution of such rewards are considered as actions.

We model the basic incentive model used by most blockchain systems such as Bitcoin. Our model provides a reward for generating a valid block (block reward) and a reward for all transactions included in a block (transaction fee). The block reward is modeled as a fixed amount of cryptocurrency that can be configured and changed by the end-user. The transaction fee is calculated as the multiplication of its size and its prize, where the prize is the amount of money the submitter of the transaction is willing to pay per unit of size. The size and the prize for transactions can also be configured as fixed or variable (random) values. However, it is possible to extend the current model to include different rewards (e.g., rewards for uncle blocks) or change the way how the fee for transactions is calculated. We model the distribution of rewards by increasing the balance of each miner after having a valid block attached to the ledger.

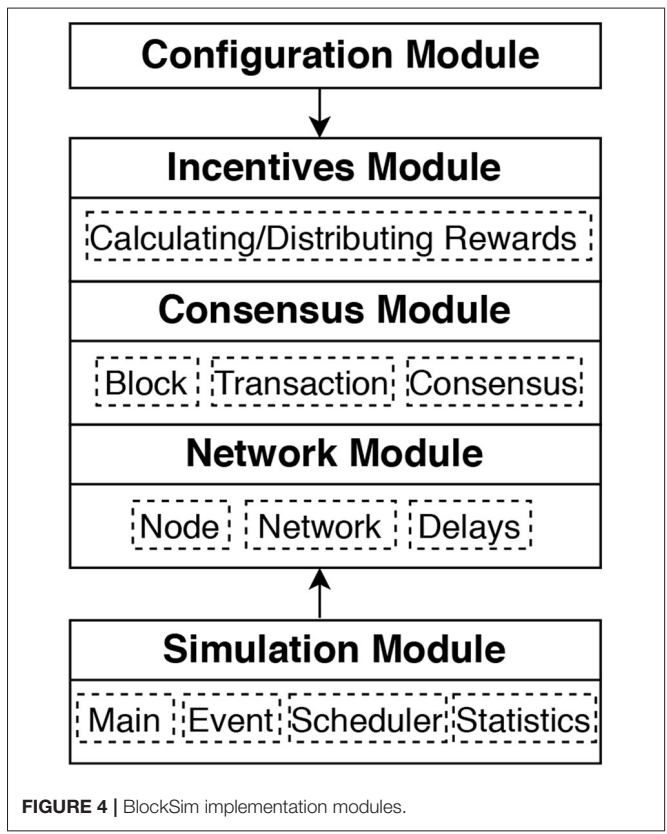

**FIGURE 4** | BlockSim implementation modules.

## 4. BLOCKSIM IMPLEMENTATION

We present the implementation of the BlockSim simulator using Python 3.6.4[1]. The main modules are given in **Figure 4**. The Simulator Module implements the core engine of the simulator, in particular the event scheduler, which we explain in section 4.1. The main topic of discussion in that section is the granularity at which events are handled, since it heavily impacts the performance of the simulator. This simulation engine module is complemented by the Configuration Module, to be described in section 4.2, which provides the user with ways to configure the simulation model and experiments. Section 4.3 explains the implementation of the Base Model, subdivided according to the main layers: Network Module, Consensus Module, and Incentives Module.

## 4.1. BlockSim Simulation Engine and Event Scheduler

As depicted in **Figure 4**, the main Simulation Module contains four classes, which are Event, Scheduler, Statistics and Main. We start with explaining our design choices for the event scheduling.

We provide event scheduling at two abstraction levels, the first one considers blocks as the event "unit," the second considers transactions as the event "unit." We explain the block-level events. The class *Event* defines the structure of events in our simulator. In the case of a block-level event it has four attributes: *type*, *nodeID*, *time*, and *block*. The attribute *type* indicates how to handle the event, in particular whether the event at hand is to *create a new block* or to *receive an existing block*. The *nodeID* and *time* attributes specify the node that handles the event and the time at which the event takes place. The *block* attribute contains the necessary information for the block to be handled.

[1] https://github.com/maher243/BlockSim

*Scheduler class* is responsible for scheduling future events and record them in the *Queue*. *Queue* is an array list that maintains all future events, and it is continuously updated during the simulation by either inserting new events or removing existing ones. At the block-level, for instance, once a block is created through a *block creation* event, the *Scheduler* class schedules *block reception* events for other nodes to receive the block. Also, it schedules a new *block creation* event by selecting a miner to propose and generate a new block on top of the last one.

The function of the *Main* and *Statistics* classes is as one would expect. *Main* runs the simulator. It prepares the setup and then triggers the *Scheduler* class to schedule some initial events. The setup includes the creation of transactions as well as the creation of the first (genesis) block, an empty block that will be attached to the local blockchain for all the nodes in the network. Then, it keeps going through all the events and executes them one by one until the *Queue* is empty or the pre-specified simulation time is reached. *Statistics* maintains the results and calculates the statistics of the final output of the simulation, including block statistics (number of blocks included in the ledger and percentage of discarded blocks), throughput and mining profits.

## 4.2. Configuration Module

This module serves as the main user interface, in which users can select from the available models as well as configuring various parameters related to the participating nodes, blocks, transactions, consensus, incentives and the simulation setups. **Table 1** summarizes the input parameters to be configured before running the simulator. We can, for instance, configure the number of nodes, the block interval time, the volume of transactions to be created per second and other parameters. Besides, our simulator allows disabling transactions if they are not of interest. This can be done by only setting the parameter *hasTrans* to be "False," without modifying the code of the simulator. Furthermore, it allows selecting a suitable technique (either full or light) for modeling transactions. If we extend the simulator by, for example, including new consensus protocols, this would be reflected in this module to allow the user of the simulator to choose the desired protocol.

## 4.3. Base Model Modules

We discuss the implementation of simulation classes that represent the Base Model of section 3 using the same three layers as before.

**Network Module:** We implement the network module in two different classes, namely, Node and Network. *Node class* defines the structure of nodes in our simulator. We implement each node as an object in which each node is given a unique ID and a balance. For each node, we assign two array lists to model the local blockchain and the transactions pool. It is worth noting that each node maintains a transactions' pool only if the full transaction technique is applied. Otherwise, a common pool will be shared by all the nodes. *Network class* implements the network latency for propagating both blocks and transactions between the nodes. Currently, we implement the latency as a time delay that can be configured by the user of the simulator in the configuration module. Hence, it could be possible to extend this class to implement a particular broadcast protocol.

TABLE 1 | Input parameters for the simulator.

| Type | Parameter | Description |
| --- | --- | --- |
| Blocks | $B_{interval}$ | Average time to generate a block in seconds |
| | $B_{size}$ | Block size in Megabyte (MB) |
| | $B_{delay}$ | Propagation delay of blocks in seconds |
| | $B_{reward}$ | Block generation reward |
| Transactions | hasTrans | Enable/Disabled transactions |
| | $T_{technique}$ | Technique for modeling transactions |
| | $T_n$ | Rate at which transactions can be created |
| | $T_{delay}$ | Propagation delay of transactions in seconds |
| | $T_{fee}$ | Transaction fee |
| | $T_{size}$ | Transaction size in MB |
| Nodes | $N_n$ | Total number of nodes in the network |
| Simulation | $Sim_{time}$ | Length of the simulation time |
| | Runs | Number of simulation runs |

**Consensus Module:** We implement the consensus module in different classes, namely, Transaction, Block and Consensus. *Transaction class* defines the structure of transactions in our simulator. We implement each transaction as an object that has seven attributes, namely, ID, timestamp, submitter ID, recipient ID, value, size, and fee. The end-user can set the size and fee of transactions in the configuration module as fixed values or random values drawn from general distributions, including exponential distribution. This class also implements both full and light techniques for modeling transactions, as we discussed in section 3.3.1. *Block class* defines the structure of blocks in our simulator. We implement each block as an object that has seven attributes, namely, depth, ID, previous ID, timestamp, size, miner ID, and transactions. This class also implements the processes required by the nodes to generate and receive blocks, as discussed in section 3.3.2. *Consensus class* implements the consensus algorithm as well as the fork resolution rule. It also implements the process of selecting leaders, aka miners, to generate and append new blocks to the ledger. This class is structured to be easy to implement any consensus protocol of interest. For instance, to implement PoW algorithm with the longest-chain rule to resolve potential forks as the case in Bitcoin and Ethereum.

**Incentives Module:** This module is responsible for setting the rewarded elements as well as calculating the rewards. Also, it distributes the rewards among the participating nodes by increasing the balance of each node after calculating the rewards. It is, however, possible to extend this module by adding more rewarded elements or changing the way the awards are calculated if required. To make it easier for the end-user, the rewards (e.g., block rewards) can be configured and changed in the configuration module.

## 5. BLOCKSIM CASE STUDIES

BlockSim is designed to be used for any type of blockchain, and to demonstrate this we apply the Base Model of BlockSim to simulate Bitcoin as well as Ethereum. We also discuss how to extend the BlockSim implementation of the Base Model to support any consensus algorithm of interest.

### 5.1. Bitcoin in BlockSim

To simulate Bitcoin we introduce the following modifications and extensions to the core implementation of BlockSim discussed in section 4.

**Network Layer:** For Bitcoin we abstract the underlying broadcast protocol by modeling the propagation of transactions and blocks as a time delay, as indicated in section 3.2. To parameterize the model one can use DSN Bitcoin Monitoring to obtain the propagation delay of information. The Node module is extended with an attribute for a node's hash power, which we add to the configuration module for the user to set as an input parameter. To distinguish between regular nodes and miners, we can assign zero as the hash power for regular nodes to indicate that the node cannot build blocks (only create and propagate transactions).

**Consensus Layer:** Bitcoin uses PoW with the longest-chain rule to resolve the forks. As discussed in section 2.2.2, in PoW miners compete against each other to be allowed to create the next block. They repeatedly draw a random number, combine it with info from the new block and generate a hash. If the hash fulfills some property, the block can be added to the blockchain and forwarded to other nodes. That means miners execute what amounts to a Bernouilli trial and since the number of trials is high, the Bernouilli trials process converges to its continuous-time counterpart, the Poisson Process. That is, the time between successes is exponentially distributed. In the configuration module, one can set the block difficulty through the $B_{interval}$ parameter, which is the time interval (in seconds) between two consecutive blocks. If multiple chains have the same depth, Bitcoin uses the longest chain to reach a global view of the blockchain ledger by resolving the forks.

**Incentives Layer:** The incentives in Bitcoin for generating blocks and executing transactions is the same as that of the Base Model. In our main BlockSim implementation, all rewards will be distributed to miners at the end of each simulation run. If needed, the Incentives module can be modified to distribute rewards in run-time. The miner of a block that is finalized and is part of the longest chain receives the block reward and the fees for all transactions included in that block. The rewards can be set in the configuration module.

### 5.2. Ethereum in BlockSim

Ethereum is very similar to Bitcoin but introduces a few additional elements associated with the handling of uncle blocks as well as attributes required for incentives associated with smart contracts.

**Network and Consensus Layers:** Ethereum allows attaching uncle blocks to a valid block and rewards miners for this. Therefore, we extend the Bitcoin Node module with an unclechain attribute. The unclechain for a node is modeled as an array list storing all chains with uncle blocks that occur during the simulation run. Ethereum allows miners to include a maximum of 2 uncle blocks within the last seven block generations (e.g., an uncle block with a depth 10 can be referenced in a block with a depth less than or equal to 17). We include this logic in the configuration module and allow configuring the maximum number of uncle blocks per block, the number of generations in which an uncle block can be included as well as disabling uncle inclusion mechanism if it is not of interest.

Similarly, we extend the Node module when receiving a block. If the block has a smaller depth or index, the block is appended to the recipient's unclechain as an uncle block to be referenced in a future block. Also, when receiving and appending a valid block to the local blockchain ledger, the miner updates its local unclechain, where necessary, by removing all the uncle blocks that have already been included in the received block.

**Incentives Layer:** The incentive model of Ethereum, similar to that of Bitcoin, includes block reward and transactions fee. Yet, Ethereum uses the Gas mechanism to calculate the fee for transactions with smart contracts. To determine the fee for transactions and blocks, we therefore require some additional attributes related to the gas model. For transactions, we add Gas Limit, Used Gas and Gas Price attributes. For blocks, we include the attributes of Gas Limit and Used Gas. We refer to the literature, e.g., Wood, 2014; Alharby and Van Moorsel, 2017 for details, but in short, Used Gas multiplied by the Gas Price corresponds to the fee the miner receives, where Used Gas depends on the computational requirements of the smart contract (Aldweesh et al., 2018), but never exceeds Gas Limit.

Ethereum also introduces rewards for uncle blocks. The uncle reward is distributed between the miner who generated the uncle and the miner who included it in his block, as follows (Wood, 2014). The miner who generated the uncle gets a variable reward depending on when the uncle has been referenced in a main block. The sooner the uncle is referenced in a block, the higher the uncle reward ($R_{uncle}$):

$$R_{uncle} = (D_{uncle} + (G_{uncle} + 1) - D_{block}) * \frac{R_{block}}{G_{uncle} + 1} \quad (1)$$

where $D_{uncle}$ is the depth of the uncle, $G_{uncle}$ is the number of generations in which the uncle can be included, $D_{block}$ is the depth of the block and $R_{block}$ is the block reward. The miner who included the uncle in his block will get a fixed reward, which is calculated as $\frac{1}{32} * R_{block}$. All this is implemented in the incentives module, but the amount of rewards can be set in the configuration module, if required.

**TABLE 2** | Data gathered from Bitcoin and Ethereum, serves as input to the simulation runs used as validation.

| Parameters | Bitcoin | Ethereum |
|---|---|---|
| $B_{interval}$ | 596s | 12.42s |
| $B_{delay}$ | 0.42s | 2.3s |
| $B_{size}$ | 0.83MB | 7,997,148Gas |
| $T_{size}$ | 546Byte | Distribution |

## 5.3. Different Consensus Protocols in BlockSim

Thus far we have mainly considered PoW as consensus protocol, but there are many other, including Proof of Stake (PoS), Proof of Authority, or message-based consensus algorithms such as Practical Byzantine Fault Tolerance and its many variants (Angelis et al., 2018).

A significant difference between these protocols and PoW is that in PoW miners are not directly selected by the consensus protocol, but instead, miners continuously invest their computing power to create the subsequent blocks. In PoS, for instance, miners would be selected by the protocol based on the amount of stake or cryptocurrencies they hold. The more cryptocurrencies a miner deposited in the system, the more chance they would be selected to generate the next block. Other protocols select miners in a round-robin manner such as Tendermint (Kwon, 2014) or based on different metrics (Angelis et al., 2018).

To support approaches such as PoS, we modify the consensus class by changing how miners are being selected to generate the next blocks. Other consensus elements (e.g., transactions, blocks, and fork resolution) and modules (simulation, network, and incentives) remain unchanged. In general, as long as the output metrics can be truthfully simulated with events scheduled at the granularity of blocks, BlockSim can be extended in a natural matter. The time consumed by the consensus algorithm would then be represented by a delay. However, if one wants to analyze the impact of specific message sequences on the performance of PBFT style consensus protocols, BlockSim is a less obvious candidate. For efficient (i.e., fast) simulation, one would study such consensus protocols through simulation tools that operate at message-level and not mix different levels of abstractions and time granularity.

## 6. BLOCKSIM VALIDATION

A nice feature of the blockchain design is that it offers invariants (such as the block creation interval) and plenty of publicly available data to validate the results of any simulator. First we compare BlockSim with existing blockchain systems (section 6.1), then we compare with various peer-reviewed studies (section 6.2).

### 6.1. Comparison With Measurements

We compare the results from BlockSim with the most popular public blockchains, Bitcoin and Ethereum. These provide certain

TABLE 3 | Validation of the simulator results by comparison with measurements from Bitcoin and Ethereum.

| | Bitcoin | | Ethereum | |
| --- | --- | --- | --- | --- |
| | Measured | Simulated | Measured | Simulated |
| $B_{included}$ | 146 ± 4 | 143 ± 5 | 6083 ± 27 | 6079 ± 25 |
| Stale (uncle) rate | 0.025% ± 0.051% | 0.049% ± 0.069% | 12.56% ± 0.43% | 12.55% ± 0.14% |
| Throughput | 2.69 ± 0.09 | 2.66 ± 0.09 | 5.99 ± 0.18 | 6.96 ± 0.03 |

$B_{included}$ is the number of blocks included in the main blockchain per day, the stale (or uncle) rate per day are blocks not in the main chain, and throughput is the number of transactions processed per second.

"invariants" that we know to be true, such as the frequency of generating blocks and the proportionality between the miner's hashing share and the probability to win the Proof of Work competition. Bitcoin and Ethereum also provide ample public data to validate our simulator.

**Validation of block and transaction metrics:** We use the following metrics for validation: number of blocks created, number of uncle or stale blocks (blocks that will not be part of the final chain), and the number of transactions completed per time unit. The results obtained from our simulator and that from the actual systems are reported in **Table 3**. We report both the average and the 95% confidence interval values, for a run of the simulation that corresponds to a full month of real time. From **Table 3**, we see that our simulator's confidence interval contains the result from the measurements. However, our simulator shows a slightly higher throughput for Ethereum compared to the real data observed. We believe that this is either due to the small sample of transactions retrieved or the fitted frequency distribution.

To obtain the above results, **Table 2** shows the data gathered from both Bitcoin and Ethereum used as input to the validation runs. That is, we use the values from **Table 2** for the relevant input parameters given in **Table 1**. We gather the Bitcoin's data from blockchain.info[2], while the Ethereum's data comes from etherscan.io[3]. We collect 1 month of data for each system as of October 2018. From these sources, we were able to directly collect all the necessary data, apart from the block propagation delay and the transactions' size in Ethereum. However, we obtain the block delay using DSN Bitcoin Monitoring[4] and ETHstats[5]. To obtain the size of transactions in Ethereum, we implement a python script that makes use of etherscan.io APIs to retrieve transactions information. We retrieve the data for the latest 5,000 transactions and then fit a frequency distribution for transactions' size to be used as input in our simulator. For the sake of this experiment, we fit a frequency distribution with the limited collected data.

**Validation of poW:** An invariant we can use for validation is the share of blocks each miner generates since it is known that share is equal to the miner's share of the overall hashing power. For instance, if a miner controls 40% of the network's hash power, it should generate 40% of the total blocks. To validate

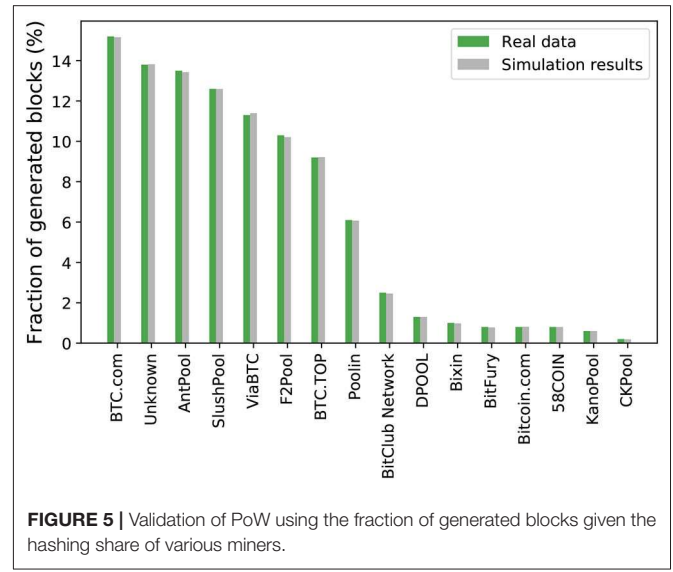

FIGURE 5 | Validation of PoW using the fraction of generated blocks given the hashing share of various miners.

PoW, we collect the estimated hash power as well as the fraction of blocks contributed by Bitcoin miners and miner pools from blockchain.info and input this into our simulator. That is, the simulation is with miners that have the same share of the hashing power as various existing Bitcoin miners.

**Figure 5** shows the results. We simulate 4 days of the Bitcoin network, a total of 1000 times and obtain the average fraction of blocks generated by each miner. The *x*-axis of **Figure 5** shows the name of the miners and the *y*-axis shows the fraction of blocks contributed by the miners for both the real Bitcoin network (the green bars) and the simulation results (the gray bars). From **Figure 5**, we see that the simulation results are very close to that of the real Bitcoin network.

## 6.2. Comparison With Peer-Reviewed Studies

We also compare the simulator results for the stale rate with that of previous peer-reviewed studies. Decker and Wattenhofer (2013) run an experiment on the Bitcoin blockchain by listening to 10,000 blocks. They found the average block propagation delay is 12.6 s and the stale rate is 1.69%. Gervais et al. (2016) run some simulation experiments using the configurations of different blockchain systems such as Bitcoin, Litecoin, and Dogecoin. They found that their simulation results matched that of the actual systems. To validate our simulator against these studies, we use the same configurations of the block interval ($B_{interval}$) and block propagation delay ($B_{delay}$) as reported in these studies. We simulate each configuration for a total of 10,000 blocks and report the average results obtained from 10 independent runs, see **Table 4**. From **Table 4**, we see that the stale rates obtained from our simulator are close to the ones reported in previous studies, with a difference of less than 10%.

---

[2]https://www.blockchain.com/explorer
[3]https://etherscan.io/
[4]https://dsn.tm.kit.edu/bitcoin/
[5]https://ethstats.net/

TABLE 4 | A comparison between BlockSim and previous studies in terms of the stale rate observed.

| | Input parameters | | Stale rate | |
|---|---|---|---|---|
| | $B_{interval}$ (s) | $B_{delay}$ (s) | Measured (%) | Simulated (%) |
| Bitcoin (Gervais et al., 2016) | 600 | 14.7 | 1.51 | 1.69±0.08 |
| Bitcoin (Decker and Wattenhofer, 2013) | 600 | 12.6 | 1.68 | 1.73±0.09 |
| Litecoin (Gervais et al., 2016) | 150 | 4.18 | 1.82 | 1.88±0.11 |
| Dogecoin (Gervais et al., 2016) | 60 | 2.08 | 2.15 | 2.38±0.08 |

## 7. BLOCKSIM SIMULATION RESULTS

To show the applicability of our simulator, we conduct a simulation experiment to investigate the impact of different consensus and network parameters on the security, performance and mining ecosystem of blockchain systems. We also show the performance of the simulator in terms of run time. We use very similar metrics as in the validation, but for a wider range of parameter values. The main discussion in this section is about how the stale block rate impacts mining decentralization and how Ethereum's approach to reward uncle blocks improves mining decentralization.

More precisely, we study the impact of different combinations of block interval and block propagation delay on the stale rate, throughput and mining decentralization. Stale rate is a security indicator of a blockchain system, and the lower the rate, the better for the security of the system (Gervais et al., 2016). Throughput represents the number of transactions that can be processed per second, thus directly indicating how well the system performs. Mining decentralization indicates that the fraction of blocks a miner includes in the main ledger is proportional to the hash power of that miner. In other words, mining decentralization means each miner gets a fair reward compared to its hash power.

Table 5 shows the results (stale rate, throughput and mining decentralization) for 25 different combinations of different block interval $B_{interval}$ and block delay $B_{delay}$ as well as the run time for every configuration. For ease of presentation, we consider only five miners (M1, M2, ..., M5) with hash powers ranging from 5 to 40%. The hash power for a miner is a configurable parameter (see section 5.1). For all configurations, we set the block size to be 1MB and the average transaction size to be 546 bytes (as in the Bitcoin network). We simulate each configuration for a total of 10,000 blocks and report the average results from 10 independent runs. The confidence intervals are not reported here, but are all within 10% of the average values.

**Stale rate:** From the stale rate results reported in **Table 5**, we observe the following. First, reducing the block interval, i.e., the time between successive blocks being created, leads to higher stale rates, especially when the block interval is already small. For instance, reducing the block interval from 12 to 1 s in the case of 0.5 s block delay will result in an increase of the stale rate by about sevenfold. When the block interval is small, other nodes could manage to find the next block before hearing of other competitive blocks due to the network latency, leading to conflicts. Also, increasing the block propagation delay leads to higher stale rates. For instance, the stale rate increases about tenfold when increasing the delay from 0.5 to 16 s in the case of 12 s block interval. The block delay includes the block's transmission time as well as the verification of the block and its embedded transactions (Decker and Wattenhofer, 2013). Thus, the bigger the block size, the more time required to transmit and verify the block. Hence, increasing the block size will result in higher stale rates. Furthermore, to ensure the lowest stale rate the block delay should be as small as possible and the block interval as large as possible. For instance, in the case of 600 s block interval, the stale rates are minimal since the block delay is only a tiny fraction of the block interval.

**Throughput:** From the throughput results reported in **Table 5**, we observe the following. First, reducing the block interval leads to higher throughput. This is because more blocks will be generated, and thus, more transactions will be processed. We also observe that the block delay could reduce the throughput significantly, especially when the block interval is small. The number of transactions that can be processed per second is reduced from 147 to 92 when increasing the block delay from 0.5 to 16 s in the case of 12 s block interval. Furthermore, the block delay does not have a significant impact on the throughput if the delay is too small compared to the block interval. For instance, in the case of 600 s block interval, the throughput achieved is almost the same even when the block delay is increased from 0.5 to 16 s.

**Mining decentralization:** From the mining decentralization results reported in **Table 5**, we observe the following. First and most importantly, we observe a correlation between stale rates and mining decentralization. The smaller the stale rates the better the mining decentralization and vice versa. In the discussion about stale rates, we observe that reducing the block interval or increasing the block delay can lead to a higher stale rate. That is, reducing the block interval leads to poor mining decentralization. In the case of 1 s block interval, for instance, miners with a large hash power (e.g., M1) have a higher fraction of blocks included in the main ledger, and thus gain higher profit, compared to their hash power invested. On the contrary, small miners have a small fraction of blocks included in the ledger, and thus gain less profit, compared to their hash power invested. Similarly, increasing the block delay negatively impacts the decentralization of the mining process. For a better mining decentralization, the stale rate should be reduced by having the block interval relatively larger than the block delay.

TABLE 5 | The simulation results (stale rate, throughput and the fraction of blocks contributed by each miner) as well as the run time performance (in seconds) for different combinations of block interval and block propagation delay.

| Input parameters | | Stale rate (%) | Throughput | Mining decentralization (% blocks contributed by miners) | | | | | Run time |
|---|---|---|---|---|---|---|---|---|---|
| $B_{interval}$ | $B_{delay}$ | | | M1(40%) | M2(30%) | M3(15%) | M4(10%) | M5(5%) | |
| 1 | 0.5 | 24.74 | 1387.74 | 43.54 | 30.13 | 13.51 | 8.65 | 4.17 | 15.9 |
| | 2 | 45.43 | 997.34 | 49.42 | 30.14 | 10.98 | 6.501 | 2.96 | 16.25 |
| | 4 | 54.81 | 829.17 | 54.76 | 29.33 | 8.96 | 4.86 | 2.09 | 13.25 |
| | 8 | 62.69 | 680.77 | 60.05 | 28.49 | 6.71 | 3.53 | 1.22 | 10.51 |
| | 16 | 68.77 | 569.93 | 66.36 | 26.48 | 4.50 | 2.0 | 0.67 | 8.29 |
| 12 | 0.5 | 3.49 | 147.26 | 40.27 | 30.15 | 14.82 | 9.86 | 4.90 | 4.41 |
| | 2 | 11.62 | 135.11 | 41.23 | 30.24 | 14.40 | 9.41 | 4.73 | 11.19 |
| | 4 | 19.33 | 123.02 | 42.60 | 30.17 | 13.77 | 9.09 | 4.37 | 14.08 |
| | 8 | 28.8 | 108.61 | 44.42 | 30.10 | 13.05 | 8.34 | 4.08 | 16.68 |
| | 16 | 39.55 | 92.09 | 47.32 | 30.39 | 11.83 | 7.30 | 3.16 | 17.64 |
| 60 | 0.5 | 0.7 | 30.17 | 40.06 | 30.05 | 14.88 | 10.15 | 4.86 | 2.21 |
| | 2 | 2.79 | 29.79 | 40.09 | 30.07 | 15.02 | 9.88 | 4.95 | 3.87 |
| | 4 | 5.36 | 28.6 | 40.50 | 29.94 | 14.94 | 9.72 | 4.90 | 5.93 |
| | 8 | 9.7 | 27.56 | 41.23 | 30.03 | 14.40 | 9.60 | 4.74 | 10.06 |
| | 16 | 16.45 | 25.51 | 42.00 | 30.24 | 14.15 | 9.11 | 4.51 | 13.12 |
| 150 | 0.5 | 0.29 | 12.23 | 39.95 | 29.94 | 15.07 | 10.03 | 5.00 | 1.86 |
| | 2 | 1.21 | 12.06 | 39.81 | 30.07 | 15.21 | 9.89 | 5.02 | 2.86 |
| | 4 | 2.23 | 11.96 | 40.32 | 30.06 | 14.76 | 9.93 | 4.94 | 3.5 |
| | 8 | 4.33 | 11.62 | 40.48 | 29.96 | 14.86 | 9.80 | 4.90 | 5.11 |
| | 16 | 8.15 | 11.3 | 40.89 | 29.78 | 14.82 | 9.77 | 4.75 | 7.95 |
| 600 | 0.5 | 0.08 | 3.03 | 39.97 | 29.98 | 15.03 | 9.96 | 5.06 | 1.67 |
| | 2 | 0.32 | 3.05 | 40.00 | 29.90 | 15.11 | 10.00 | 4.99 | 1.84 |
| | 4 | 0.61 | 3.05 | 40.10 | 30.24 | 14.83 | 9.85 | 4.98 | 2.23 |
| | 8 | 1.14 | 3.01 | 39.91 | 29.99 | 15.17 | 9.91 | 5.02 | 2.91 |
| | 16 | 2.26 | 2.99 | 39.98 | 29.99 | 15.07 | 10.04 | 4.92 | 3.67 |

**Run time performance:** For every combination of configurations, we show the average time (in seconds) it takes the simulator to perform a single run. To obtain the run time results, we use a laptop with a 2.30GHz Intel i5 CPU with 16GB RAM running on Windows 10. From **Table 5**, we observe the following. First, the run time generally takes seconds to simulate 10,000 blocks. We note that in this experiment there are five miners and increasing the number of miners would increase the run time since more actions need to be performed in the network. For example, every new miner has to maintain a ledger and update it every time a new block is announced in the network. At the same time, increasing the number of non-miners would not affect the run time that much as they are not participating in maintaining the ledger. Secondly, the run time increases for higher stale rates (setting with small $B_{interval}$ or large $B_{delay}$). This is because miners need to update their ledgers more frequently than when conflicts are rare. Surprisingly, when the stale rate is high (over 50%) the run time seems to be decreasing. We believe the explanation for this is that although more blocks are in the system, miners neglect most blocks as they arrive when the miner is behind the main chain.

**Bitcoin throughput:** The current implementation of Bitcoin compromises of 596 s block interval and 0.42 s block delay, as reported in **Table 2**. That means the Bitcoin network experiences a low stale rate as well as a good mining decentralization. However, it suffers from poor throughput as the number of transactions processed per second is only about 3. We argue that we could securely reduce the block interval of Bitcoin to 60 s to improve the throughput by about a factor 10, without any significant impact on the stale rate or mining decentralization.

**Ethereum mining decentralization through uncle inclusion:** The current implementation of Ethereum compromises of 12.42 s block interval and 2.3 s block delay, as reported in **Table 2**. This results in a stale rate of about 12.56% and imperfect mining decentralization, but a better throughput than the Bitcoin blockchain. To eliminate the negative impact on the stale rate and mining decentralization, Ethereum uses an uncle inclusion mechanism, where stale blocks are included in the main ledger as uncle blocks and the miners of such blocks are rewarded. However, this does not guarantee that miners will receive fair rewards compared to their hash power invested (e.g., a miner with a hash power of 20% should receive 20% of the total rewards distributed in the network). This is especially true as miners get a lower reward for uncle blocks compared to main blocks as well as they are not rewarded for the transactions included in the uncle blocks.

TABLE 6 | The fraction of rewards gained by each miner (M1,M2,…,M5), with and without uncle inclusion mechanism.

| Miners (%) | | Fraction of rewards | |
|---|---|---|---|
| | | Without uncle inclusion (%) | With uncle inclusion (%) |
| M1 | 40 | 41.32 | 40.2 |
| M2 | 30 | 30.28 | 30.18 |
| M3 | 15 | 14.47 | 14.91 |
| M4 | 10 | 9.34 | 9.85 |
| M5 | 5 | 4.6 | 4.86 |

We use the same parameters as currently in Ethereum to further explore whether the fraction of rewards a miner would receive with uncle inclusion mechanism is proportional to its hash power. We execute 10 independent simulation runs of 10,000 blocks and report the average results in **Table 6**. From **Table 6**, we see that the fraction of rewards gained by the miners with uncle inclusion mechanism is closer to their hash power than in the case where the uncle mechanism is not applied. Thus, Ethereum indeed achieves a better mining decentralization using its uncle inclusion mechanism.

## 8. DISCUSSION: EVALUATION OF BLOCKSIM AGAINST DESIGN OBJECTIVES

We evaluate our simulator against the design criteria mentioned in section 3.1, which are generality, extensibility and simplicity.

**Generality:** Generality refers to the ability to use BlockSim for a variety of analysis questions and for a variety of blockchains. The key technology to achieve generality is the BlockSim Base Model, which has been designed in such a way that many blockchain systems and analysis questions can be answered. The Base Model covers all common building blocks of blockchains such as nodes, transactions, blocks, blockchain ledger, fork resolution and incentive models. We have demonstrated the application of blockchain to analyze Bitcoin and Ethereum, and arguably BlockSim is well-suited for the full class of permissionless blockchain systems. Furthermore, BlockSim achieves generality by supporting different properties and metrics such as performance (both throughput and latency), functionality metrics such as stale rates and system properties such as mining decentralization and mining incentives. To further support this criterion, however, we aim to model and implement different consensus protocols (e.g., Proof-of-Stake) as well as different generic broadcast protocols for the Network layer in a later version of BlockSim.

**Extensibility:** Extensibility refers to the ability of the BlockSim tool to be extended in a natural manner for various systems and analysis problems. This comes down to the design of the software, which is through modules that can easily be manipulated and extended to investigate different properties or problems of interest. The user of the simulator can use common object oriented programming techniques such as inheritance to extend current modules either by adding new functionalities (classes, methods or attributes) or modifying (overriding) some of the existing ones.

In sections 5.1, 5.2, we show how we extend the base modules of BlockSim to support the implementation of Bitcoin and Ethereum. For instance, we extend the Node module by adding an attribute for a node's hash power.

As another example, we will briefly explain how to extend BlockSim to support different malicious behaviors of the nodes (e.g., selfish mining strategies). The current implementation of BlockSim assumes that all nodes are honest. To support such behaviors, we can extend the Node module by introducing a new attribute (e.g., selfish) for each behavior. Hence, each behavior needs to be adequately defined (e.g., by writing a function or a separate class that specifies the procedures involved in this behavior).

To establish selfish mining behavior for a node, for instance, we configure that node to work on its fork without propagating the blocks it generates to other nodes in the network. Once the behaviors are defined, the user of the simulator has only to access the configuration module and choose which type of behaviors to be studied when defining the nodes, without modifying the underlying code of the simulator. Similar to this is the study of the uncertainty problem miners face during the selection of transactions (Alharby and Van Moorsel, 2018), and the analysis of the Ethereum Verifier's Dilemma (Alharby et al., 2020).

**Simplicity:** BlockSim achieves this criterion as it has been implemented in different modules as well as it provides a user interface (a configuration module) that allows the end-user to set up the input parameters for the simulator. This makes BlockSim easy to use and understand. Besides, the current version of BlockSim hides and abstracts many details. For example, it abstracts all the details of the network layer by only introducing a configurable time delay for information propagation to model this layer. Also, it hides details about the validation process of blocks and transactions. By doing so, BlockSim becomes simple and easy to use and understand. Although hiding and abstracting details can result in an incomplete model, it is possible to extend BlockSim to incorporate these details if required.

## 9. RELATED WORK

In the literature, there are some attempts to utilize simulation models to evaluate various aspects of blockchain systems. In Yasaweerasinghelage et al. (2017), the authors use architectural modeling and simulation to measure the latency in blockchain systems under different configurations. In Alharby and Van Moorsel (2018), the authors propose a simulation model to investigate the impact of profit uncertainty in the Ethereum blockchain. They found that miners in Ethereum are not able to make informed decisions about which transactions to include in their blocks to maximize their revenue. In Neudecker et al. (2015), the authors propose a simulation model to analyze and evaluate attacks on the Bitcoin network. In Göbel et al. (2016), the authors use discrete-event simulation to study the behavior of Bitcoin miners (including selfish-mining strategies) when there is a delay in propagating information among miners. Besides these proposals, there are some blockchain simulators proposed in the literature. In Gervais et al. (2016), the authors propose a Bitcoin simulator to analyze the security and performance of different configurations in both the consensus and network layers.

Several others Bitcoin-like network simulators are proposed in the literature such as Aoki et al. (2019), Miller and Jansen (2015), and Stoykov et al. (2017). However, these proposals utilize simulation-based models to study specific aspects of blockchain systems. They neither cross different layers nor cover all common functional building blocks (e.g., blocks and transactions) for blockchain systems. For instance, neither of these proposals model transactions in the blockchain system nor capture the incentives layer in the same detail as BlockSim.

With BlockSim we provide a general-purpose, widely usable, simulation tool for blockchains, to assist in answering a variety of design and deployment questions. Our discrete-event simulator generalizes on the ones proposed in the related literature by integrating different layers of the blockchain system to gain a more comprehensive insight into different aspects such as performance, security, and incentives. In BlockSim, for instance, we take a step further by considering the functional blocks common across the different implementation of blockchain systems. We design and structure BlockSim to cross different layers of blockchains. Furthermore, we model transactions in two different ways, each of which for specific purposes as well as modeling both Bitcoin and Ethereum blockchains.

## 10. CONCLUSION

This paper proposes BlockSim, a discrete-event simulation framework for blockchain systems, capturing network, consensus and incentives layers of blockchain systems. The simulation tool is implemented in Python and is available for general use. We introduce the design and evaluate it against the design objectives of generality, extensibility and simplicity.

BlockSim's results have been validated by comparing it with design properties and measurement studies available from real-life blockchains such as Bitcoin and Ethereum. We also demonstrated the use of BlockSim in a study of stale rate, throughput and mining decentralization across a variety of blockchain configurations.

Future work should further demonstrate the extensibility of BlockSim by implementing additional variants of blockchain systems, such as those based on Proof of Stake as well as blockchains augmented with channels. In addition, one can build on the current version of BlockSim and extend it with additional reusable classes that represent other important system aspects and mechanisms, in particular mining pools and channels.

## DATA AVAILABILITY STATEMENT

The datasets generated for this study are available on request to the corresponding author.

## AUTHOR CONTRIBUTIONS

MA has design, implement, evaluate the simulation tool. In addition, he writes the whole paper. AM helps during all the processes by giving feedbacks and improving the writing of the paper.